\documentclass[]{zakoproc}
\usepackage{graphicx}
\usepackage{wrapfig}
\begin{document}

\title{Modeling Poynting flux vs. kinetic-energy dominated~jets}
\author{Mart\'{\i}n Huarte-Espinosa\footnote{martinhe@pas.rochester.edu},
Adam Frank and Eric G. Blackman}
\institute{Department of Physics and Astronomy, University of Rochester,
New York, USA.}
\markboth{M. Huarte-Espinosa}{Poynting flux vs. kinetic-energy dominated jets}

\maketitle
\vspace{-25pt}
\begin{abstract}
We present 3D-MHD AMR simulations of Poynting flux dominated (PFD)
jets formed by injection of magnetic energy. We compare their
evolution with a hydrodynamic jet which is formed by injecting kinetic
energy with the same energy flux than the PFD jets.  We predict
characteristic emission distributions for each of these jets.  
Current-driven perturbations in PFD jets are amplified
by both cooling and rotation for the regimes studied: Shocks and
thermal pressure support are weakened by cooling, making the jets
more susceptible to kinking. Rotation amplifies the toroidal magnetic
field which also exacerbates the kink instability.
\end{abstract}

\section{Introduction} 
\vskip-.3cm
Jets are observed in Young Stellar Objects,
post-AGB stars and in other astrophysical objects.  Models suggest
that jets are launched and collimated by accretion, rotation and
magnetic mechanisms (Pudritz et al.~\cite{pudritz}).  Magnetized
jets have recently been formed in laboratory experiments (Lebedev
et~al.~\cite{leb}). The importance of the magnetic fields relative
to the flows' kinetic energy divides jets into (i) Poynting flux
dominated (PFD; Shibata \& Uchida~\cite{shibata}), in which
magnetic fields dominate the jet structure, (ii)
magnetocentrifugal (Blandford \& Payne~\cite{bland}), in which
magnetic fields only dominate out to the Alfv\'en radius.  
The observable differences between PFD and 
magnetocentrifugal jets are unclear, as are the effects that cooling and 
rotation have on PFD jets.

\vspace{-10pt}
\section{Model}
\vskip-.3cm
We use the Adaptive Mesh Refinement code AstroBEAR2.0 (Cunningham
et~al.~\cite{bear}) 
to solve the equations of radiative-MHD in 3D.
The grid represents 160$\times$160$\times$400\,AU
divided into 64$\times$64$\times$80 cells plus 2 adaptive refinement
levels. Initially,
the molecular gas is static and has an ideal gas equation of state
($\gamma=\,$5$/$3), a number density of 100\,cm\,s$^{-1}$ and 
a temperature of 10000\,K.
The magnetic field is helical, centrally localized and given 
by the vector potential (in cylindrical coordinates) 
${\bf A}(r,z) =  [r/4(cos(2r) + 1)( cos(2z)
+ 1 )] \hat{\phi} + [\alpha/8 \,(cos(2r) + 1)( cos(2z) + 1 )] \hat{k}$,
for $r,z <$ 30\,AU, and ${\bf A}(r,z) =\,$0 elsewhere. 
$\alpha=\,$40, has units of length and determines the ratio of
toroidal to poloidal magnetic fluxes.
The magnetic pressure exceeds the thermal pressure
inside the magnetized region.

Source terms continually inject magnetic or kinetic energy 
at cells $r,z<$30\,AU.
We carry out 4 simulations: an adiabatic, a cooling (Dalgarno
\& McCray~\cite{dm}) and a rotating (Keplerian) PFD jet,
as well as a hydrodynamical jet.
The latter is constructed  to have same time average propagation speed and energy
flux as the adiabatic PFD jet. See Huarte-Espinosa et~al.~(\cite{we}) for
details.

\begin{wrapfigure}[22]{r}{0.57\textwidth}
\vspace{-40pt}
\begin{center}
\includegraphics[width=0.56\textwidth]{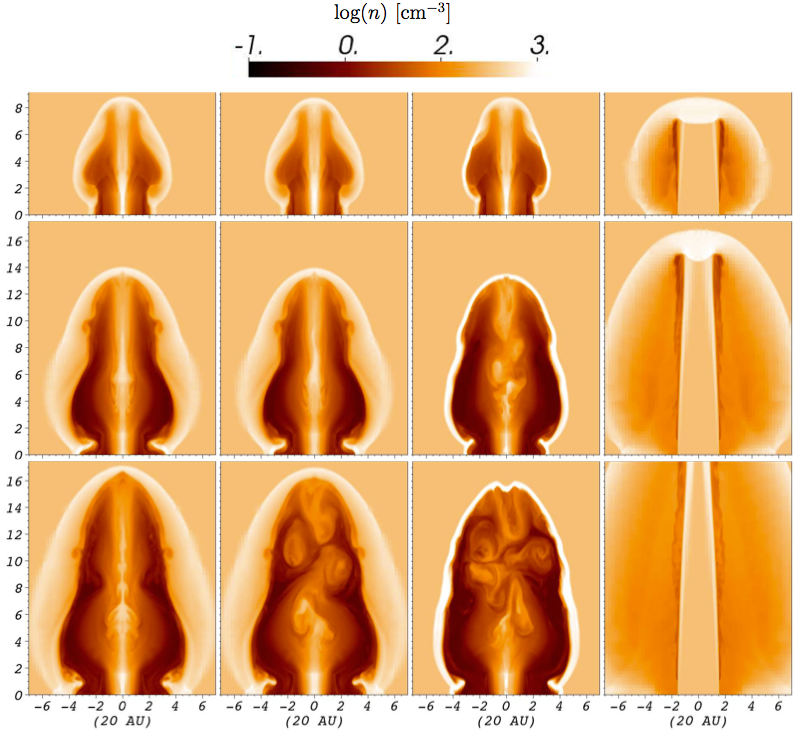} \\
\end{center}
\vspace{-20pt}
\caption{Logarithmic density maps of the adiabatic (1st column), rotating
(2nd column) and cooling (3rd column) PFD jets. Hydrodynamic jet 
(4th column). From top to bottom the time is 42, 84 and 118\,yr.}
\vspace{-10pt}
\end{wrapfigure}

\section{Results}
\vskip-.3cm
Magnetic pressure pushes field lines and plasma up, forming magnetic
cavities with low density. The adiabatic case is the most stable.
PFD jets decelerate relative to the hydro one; 
the PFD case
produces not only axial but radial expansion. 
PFD jet cores
are thin and unstable, whereas the hydro jet beam is thicker,
smoother and stable. The PFD jets are sub-Alfv\'enic. Their cores
are confined by magnetic hoop stress, while their surrounding
cavities are collimated by external thermal pressure. PFD jets carry
high axial currents which return along their outer
contact discontinuity. The PFD jets develop current-driven perturbations 
which are amplified
by cooling, firstly, and by rotation, secondly, for the regimes 
studied.

\vspace{-10pt}
\section{Conclusions}
\vskip-.3cm
PFD jet beams are lighter, slower and less stable than kinetic-energy
dominated ones.  We predict characteristic emission distributions
for each of these.  Current-driven perturbations in PFD jets are
amplified by cooling, firstly, and base rotation, secondly: Shocks
and thermal pressure support are weakened by cooling, making the
jets more susceptible to kinking. Rotation amplifies the toroidal
magnetic field which also exacerbates the kink instability.  Our
simulations agree well with the models and experiments of Shibata \&
Uchida~(\cite{shibata}) and Lebedev et~al.~(\cite{leb}), respectively.

{\tiny Acknowledgements: Financial support for this project was
provided by the Space Telescope Science Institute grants HST-AR-11251.01-A
and HST-AR-12128.01-A; by the National Science Foundation under
award AST-0807363; by the Department of Energy under award DE-SC0001063;
by NSF  grant PHY0903797,  and by Cornell University grant 41843-7012.
}

\end{document}